\documentclass[journal=apchd5,manuscript=letter]{achemso}

\usepackage[version=3]{mhchem} 
\usepackage[T1]{fontenc}       
\usepackage{siunitx}
\usepackage{lmodern}

\SectionNumbersOff

\author{Hoan Vu}
\email{hoan.vu@chemie.uni-hamburg.de}
\affiliation[Chemie]
{Institute of Physical Chemistry, University of Hamburg, Grindelallee~117, 20146~Hamburg, Germany}
\alsoaffiliation[Physik]
{Institute of Nanostructure and Solid State Physics, University of Hamburg, Jungiusstrasse~11, 20355~Hamburg, Germany}
\author{Jan Siebels}
\affiliation[Chemie]
{Institute of Physical Chemistry, University of Hamburg, Grindelallee~117, 20146~Hamburg, Germany}
\author{David Sonnenberg}
\affiliation[Physik]
{Institute of Nanostructure and Solid State Physics, University of Hamburg, Jungiusstrasse~11, 20355~Hamburg, Germany}
\author{Stefan Mendach}
\affiliation[Physik]
{Institute of Nanostructure and Solid State Physics, University of Hamburg, Jungiusstrasse~11, 20355~Hamburg, Germany}
\author{Tobias Kipp}
\affiliation[Chemie]
{Institute of Physical Chemistry, University of Hamburg, Grindelallee~117, 20146~Hamburg, Germany}

\title[Title]
{Tunable plasmonic nanoantennas in rolled-up microtubes coupled to integrated quantum wells}


\begin{document}

\begin{tocentry}
\includegraphics{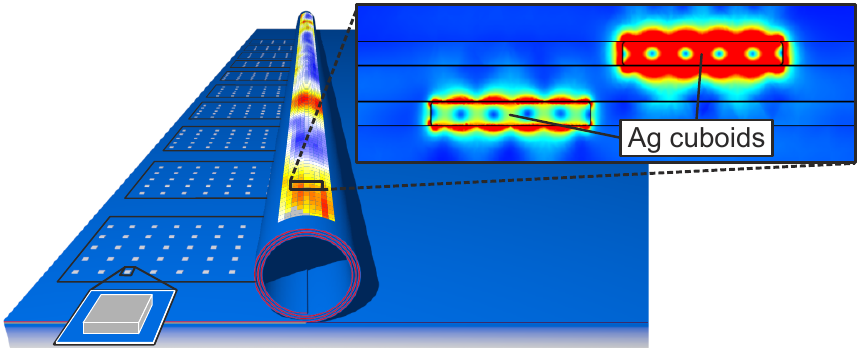}
\end{tocentry}


\begin{abstract}

We propose and realize a tunable plasmonic nanoantenna design consisting of two stacked Ag cuboids that are integrated into a rolled-up semiconductor microtube. The antenna's resonance is tuned by varying the cuboid's distance to match the photoluminescence emission of an embedded GaAs quantum well. Spatially, spectrally and temporally resolved photoluminescence measurements reveal a redshift and a reduction in lifetime of the quantum-well emission as signatures for the coupling to the antenna system. By means of finite-element electromagnetic simulations we assign the coupling to an excitation of a high-order plasmonic mode inside the Ag cuboids.

\end{abstract}


Noble metal nanostructures allow to control and manipulate light-matter interactions in the sub-wavelength regime \cite{Ozbay2006,Lal2007,Schuller2010}. Their unique optical properties are governed by localized surface plasmons (LSPs), collective charge-carrier oscillations which are accompanied by both strongly confined electromagnetic fields and large field intensities. The capability of metal nanostructures to support LSPs has been exploited to remarkably modify light-emission properties of emitters such as increasing their fluorescence intensity \cite{Song2005,Pompa2006,Kuhn2006,Taminiau2008,Tanaka2010,Kinkhabwala2009,Acuna2012,Akselrod2016}, altering their radiative and non-radiative decay rates \cite{Kuhn2006,Akselrod2016,Anger2006,Muskens2007,Ma2010} or reshaping their emission spectra \cite{LeRu2007,Ringler2008,Zhao2011}. In this regard, the nanostructures are considered as plasmonic nanoantennas and the coupling to a close-by emitter is inherently dependent on the antennas' composition, size, and geometry. Correspondingly, a manifold of antenna designs has been described including bow-tie \cite{Fromm2004,Kinkhabwala2009} and Yagi-Uda antennas\cite{Curto2010,Dregely2011}, metallic nanorods \cite{Taminiau2008,Ameling2010} and particle dimers \cite{Bakker2008,Muskens2007,Acuna2012}. The variety of designs allows to reach a wide spectral range of LSP resonances but essentially requires a pinpoint control over the antenna's geometry and a close distance of nanoantenna and emitter.

In this Letter, we propose and demonstrate the fabrication of a novel nanoantenna consisting of stacked Ag cuboids that are integrated into a rolled-up semiconductor microtube. The semiconductor compound includes a GaAs quantum-well (QW) heterostructure and thus the nanoantenna is directly placed adjacent to a robust quantum emitter. Microtubes with a functional metal layer have been previously exploited to realize rolling-up metamaterials \cite{Schwaiger2009,Rottler2012,Zhang2015,Schulz2016} and LSP-coupled whispering-gallery-mode resonators \cite{Zhang2015a,Yin2016}. Here, we combine the defined rolled-up mechanism with a precise electron beam lithography approach to stack individual Ag cuboids into the wall of the micron-sized tube. By design, the distance of the Ag cuboids is varied and allows to tune the nanoantenna's LSP resonance to match the photoluminescence (PL) emission energy of the embedded GaAs QW. We show by means of spatially and temporally resolved PL spectroscopy that the coupling of the GaAs QW emission to the plasmonic nanoantenna system is dependent on the lateral distance of the individual cuboids. For a distance of $d_x=\SI{300}{\nm}$, we observe a redshift of the QW PL emission that is accompanied by a decrease in lifetime at the red-shifted spectral position. By means of finite-element electromagnetic simulations we attribute the spectral shift to an excitation of a high-order plasmonic mode inside the Ag cuboid nanoantenna.

\begin{figure*}
		\includegraphics[scale=0.175]{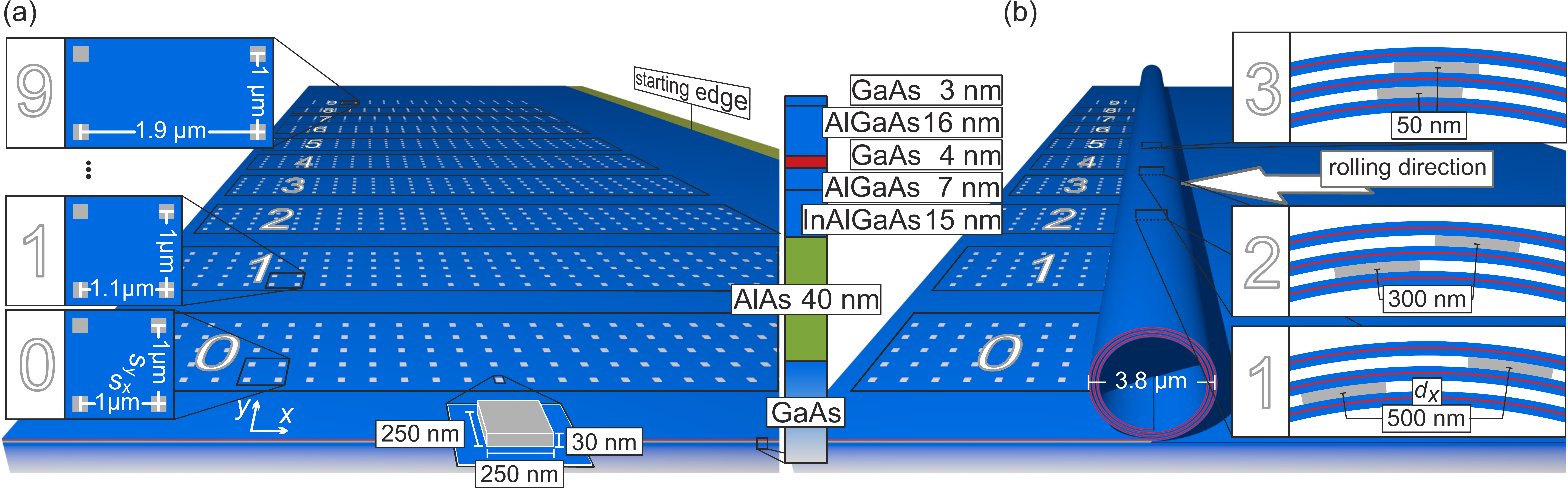}
	\caption{(a) Sketch of the sample before the rolling-up process. The inset on the right shows the molecular-beam epitaxially grown semiconductor layer structure. Identical Ag cuboids with size \SI[product-units = brackets]{250x250x30}{\nm^3} are fabricated by means of electron beam lithography and thermal layer evaporation. The Ag cuboids are grouped to arrays which are denoted with numbers $0$ to $9$. The center-to-center separation $s_x$ in rolling direction between cuboids is varying from \SIrange{1.0}{1.9}{\um} in \SI{0.1}{\um} steps for arrays $0$ to $9$. The perpendicular spacing $s_y$ is fixed to \SI{1.0}{\um} within each array. (b) Sketch of the sample after rolling up. The rotation number and the gap between Ag arrays and the starting edge are chosen such that the Ag cuboids inside the microtube are sandwiched between three semiconductor slabs, as shown in the insets. The distance $d_x$ between two cuboids is determined by their initial center-to-center separation $s_x$.}
	\label{fig:sketch}
\end{figure*}

\begin{figure}
		\includegraphics[scale=0.175]{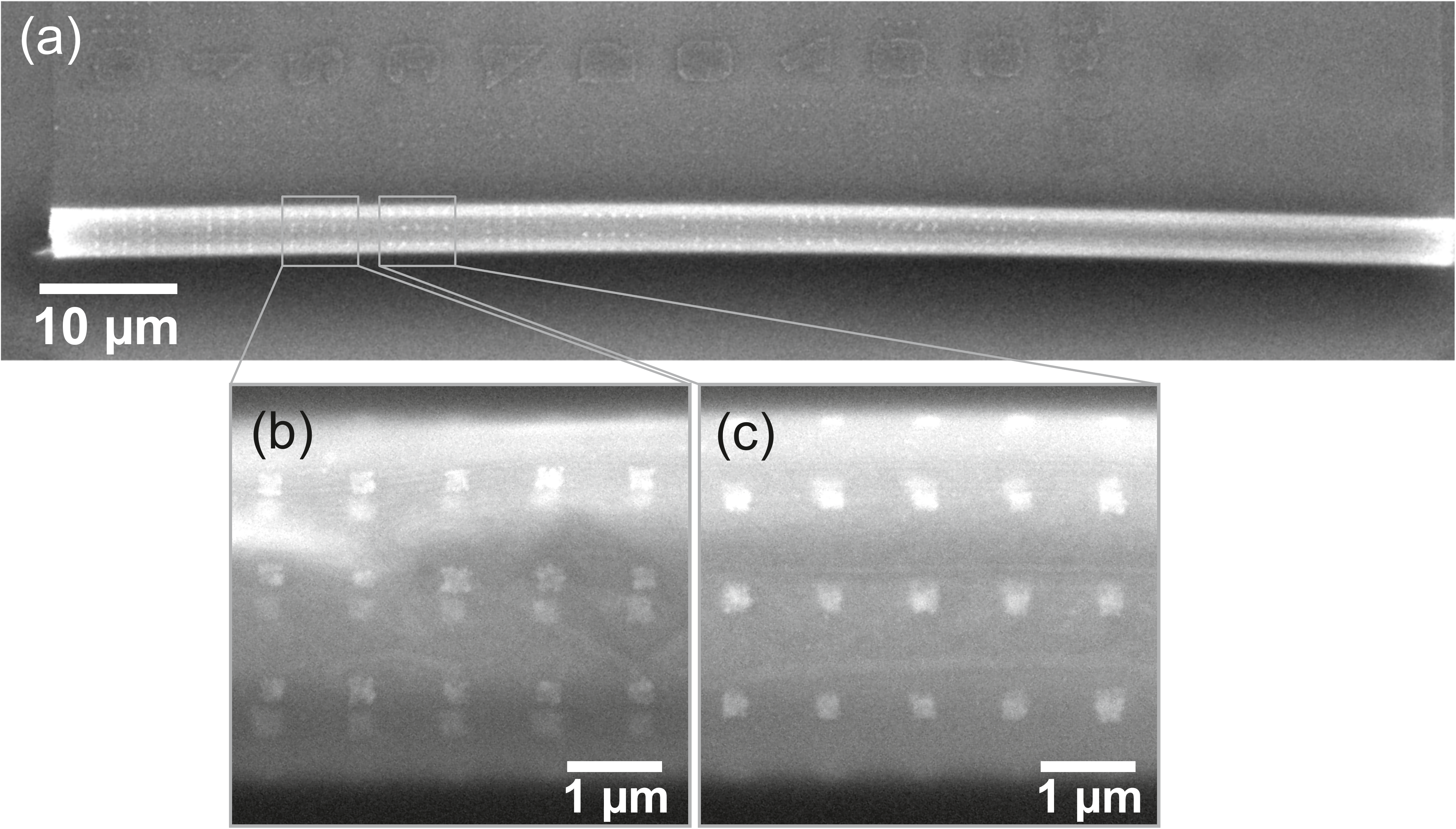}
	\caption{(a) Scanning electron micrograph of the microtube investigated in this study with \SI{100}{\um} in length and \SI{3.8}{\um} in diameter. (b,c) Magnified micrographs of Ag cuboids in array $2$ and $3$, respectively.}
	\label{fig:rem}
\end{figure}

The microtubes are fabricated by rolling-up prestrained molecular-beam epitaxially grown semiconductor layers \cite{Prinz2000,Schmidt2001}. Figure \ref{fig:sketch}\,(a) illustrates the sample layout before the rolling-up process; the detailed semiconductor structure is depicted as an inset and consists of a GaAs buffer layer, a \SI{40}{\nm} AlAs sacrificial layer, a \SI{15}{\nm} InAlGaAs strained layer, a QW heterostructure (\SI{7}{\nm} AlGaAs, \SI{4}{\nm} GaAs, \SI{16}{\nm} AlGaAs), and a \SI{3}{\nm} GaAs capping layer. Ultraviolet (UV) lithography is applied to create necessary structures for the rolling-up process, i.e., starting edges and shallow mesas (see, e.g., Ref. \cite{Mendach2006} for details) and to construct orientation markers for a following electron-beam (EB) lithography step, which allows to precisely define a resist mask consisting of cuboid arrays and its position with respect to the microtube's orientation. After the lithography steps, the resist mask is transferred to Ag cuboids by depositing a $30$-nm-thick Ag functional layer via thermal evaporation and a subsequent lift-off process. We choose cuboids of size \SI[product-units = brackets]{250x250x30}{\nm^3} as building blocks for the stacked nanoantennas. Their center-to-center separation $s_x$ in rolling direction is varied from \SIrange{1.0}{1.9}{\um} in \SI{0.1}{\um} steps for the cuboid arrays labeled with $0$ to $9$, see Fig.\,\ref{fig:sketch}\,(a). The cuboid separation $s_y$ perpendicular to the rolling direction is fixed to \SI{1.0}{\um} within each array. The rolling-up process is initiated by selective wet etching of the AlAs sacrificial layer in hydrofluoric acid. Thus, the overlying layers are released from the substrate and the strain relaxation in the InAlGaAs layer induces the self-rolling of the layers to a microtube with a typical diameter of \SI{4}{\um} and an adjustable rotation number. Figure \ref{fig:sketch}\,(b) sketches the final sample layout including a microtube with stacked Ag cuboids. The rotation number and the gap between Ag arrays and the starting edge are chosen such that two identical Ag cuboids are sandwiched between three semiconductor slabs, as depicted in the insets of Fig.\,\ref{fig:sketch}\,(b). The distances $d_x$ of the cuboids inside the microtube are determined by their initial lateral separation $s_x$. Consequently, each of the $10$ arrays corresponds to a certain distance $d_x$. For example, the cuboids of array $2$ exhibit a distance $d_x=\SI{300}{\nm}$ inside the microtube, the cuboids of array $3$ have a distance $d_x=\SI{50}{\nm}$. The distance $d_x$ strongly determines the plasmonic interaction of the Ag cuboids, their coupling, and thus their properties as a compound nanoantenna. Figure \ref{fig:rem}\,(a) shows a scanning-electron micrograph of the microtube investigated in this study with \SI{100}{\um} in length and \SI{3.8}{\um} in diameter. The cuboid arrays are visible both in their rolled-up and flat geometry. Furthermore, also lithographically defined numbers $0$ to $9$ labeling the arrays are faintly visible in the flat surface region. The magnified micrographs in Fig.\,\ref{fig:rem}\,(b) and Fig.\,\ref{fig:rem}\,(c) display the rolled-up Ag cuboids in array $2$ and $3$, respectively. Here, the cuboids appear pairwise with a different contrast owing to the fact that they are sandwiched between different layers. Such a visibility of the cuboids implies a compactly rolled microtube. This allows to directly derive the cuboid distances $d_x$ from the geometric parameters lateral separation $s_x$, distance between starting edge and arrays, individual layer thicknesses, and microtube diameter.
 
We investigate the coupling of the GaAs QWs to the Ag cuboid nanoantennas by means of spatially and temporally resolved PL spectroscopy. The experiments are carried out in a home-build laser scanning confocal microscope equipped with a closed-cycle cryostat operating at \SI{6}{\K} and with a typical spatial resolution of about \SI{500}{nm}. The QWs are pumped with circular polarized laser light at a wavelength of $\lambda=\SI{650}{\nm}$ generated by an optical parametric oscillator that is driven by a pulsed Ti:sapphire laser at a wavelength of $\lambda=\SI{830}{\nm}$. The light is focused onto the sample with a microscope objective (100x, NA = 0.8). The same objective is used to collect the emitted light which is then coupled to a streak camera system.

\begin{figure*}
		\includegraphics[scale=0.175]{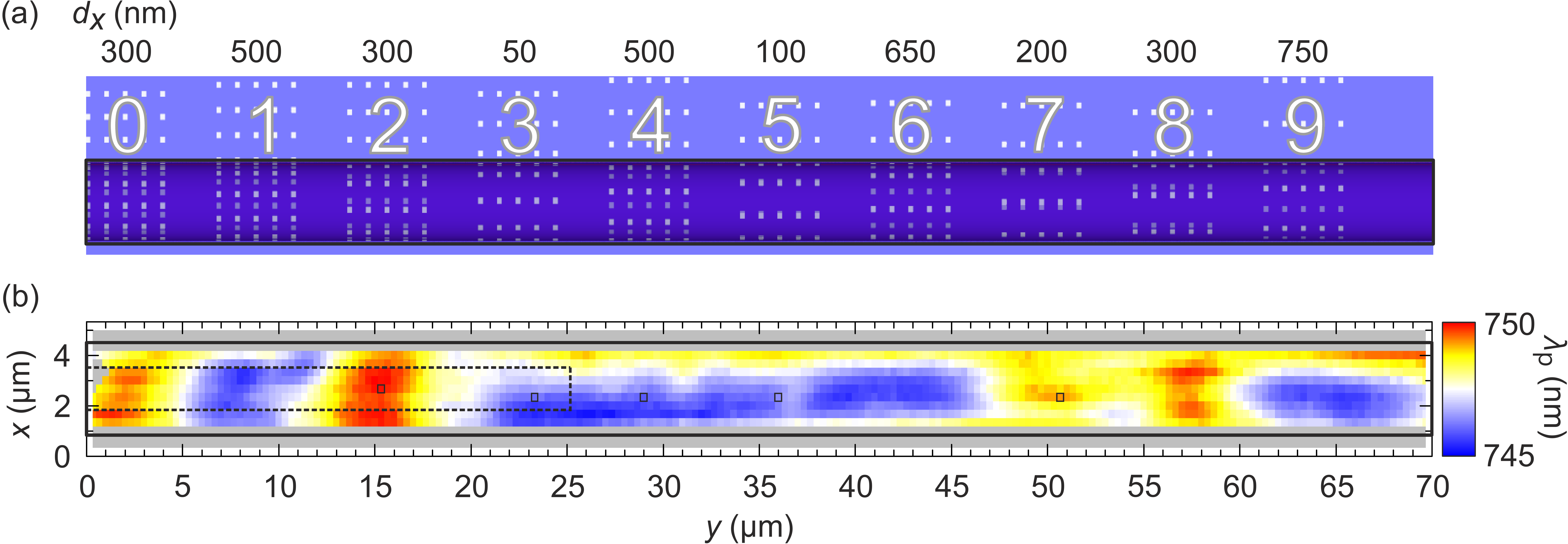}
	\caption{(a) Sketched top view of the microtube already shown in Fig.\,\ref{fig:rem}. The same label numbers as in Fig.\,\ref{fig:sketch} are used. The distance $d_x$ of the cuboids for each array is shown on the top. (b) Spatially resolved map of the wavelength position of the QW PL intensity maximum. The spectral position of the maximum PL intensity $\lambda_p$ is encoded as false color. A cross sectional profile from the area framed in dashed lines is shown in Fig.\,\ref{fig:profiles}. QW emission spectra for marked pixels are shown in Fig.\,\ref{fig:sim}\,(a).}
	\label{fig:ext_sim_data}
\end{figure*}

Figure \ref{fig:ext_sim_data}\,(a) illustrates a to-scale sketch of the sample from the top view with the microtube's contour outlined by a black frame. The same label numbers for the Ag cuboid arrays as in Fig.\,\ref{fig:sketch} are used; on the top the corresponding distances $d_x$ of the cuboids inside the microtube are shown. The distances are directly derived from the geometric parameters and are verified with SEM images with an uncertainty of about \SI{10}{nm}. We raster scanned the sample in an area of \SI[product-units = brackets]{5x70}{\um^2} with step sizes of \SI{0.3}{\um} and acquired a streak camera image at each pixel. We first evaluated the QW emission by fitting the time-integrated PL spectra with a Gaussian function. The fitted spectral position of the maximum PL intensity $\lambda_p$ is plotted in the $745$--$\SI{750}{\nm}$ wavelength range as a spatially-resolved false-color map in Fig.\,\ref{fig:ext_sim_data}\,(b). Fitted spectral positions outside this range are depicted as gray pixels. We also restricted the evaluation spatially to the area corresponding to the top-most section of the microtube; that way omitted pixels are plotted in gray as well. Both the plot frame of Fig.\,\ref{fig:ext_sim_data}\,(b) and the sketch in Fig.\,\ref{fig:ext_sim_data}\,(a) are matched to the same dimensions which allows to assign the arrays to the respective areas in Fig.\,\ref{fig:ext_sim_data}\,(b). The spatially resolved map of the wavelength position of the QW PL intensity maximum displayed in Fig.\,\ref{fig:ext_sim_data}\,(b) reveals red-shifted PL emission of the QWs in sections corresponding to the arrays $0$, $2$, $7$, and $8$. For these arrays the distances of the cuboids are $d_x=\SI{300}{\nm}$ (arrays $0$, $2$, and $8$) and $d_x=\SI{200}{\nm}$ (array $7$). The QWs in arrays with smaller distances $d_x$ (array $3$ with $d_x=\SI{50}{\nm}$, array $5$ with $d_x=\SI{100}{\nm}$) and QWs in arrays with greater distances $d_x$ (array $1$ and array $4$ with $d_x=\SI{500}{\nm}$, array $6$ with $d_x=\SI{650}{\nm}$) show no red-shifted PL emission. From the spatially and spectrally resolved measurements we therefore assume that the QW emission couples to the Ag cuboid nanoantenna for cuboid distances of $d_x=\SI{200}{\nm}$ and $d_x=\SI{300}{\nm}$. 

\begin{figure}
		\includegraphics[scale=0.175]{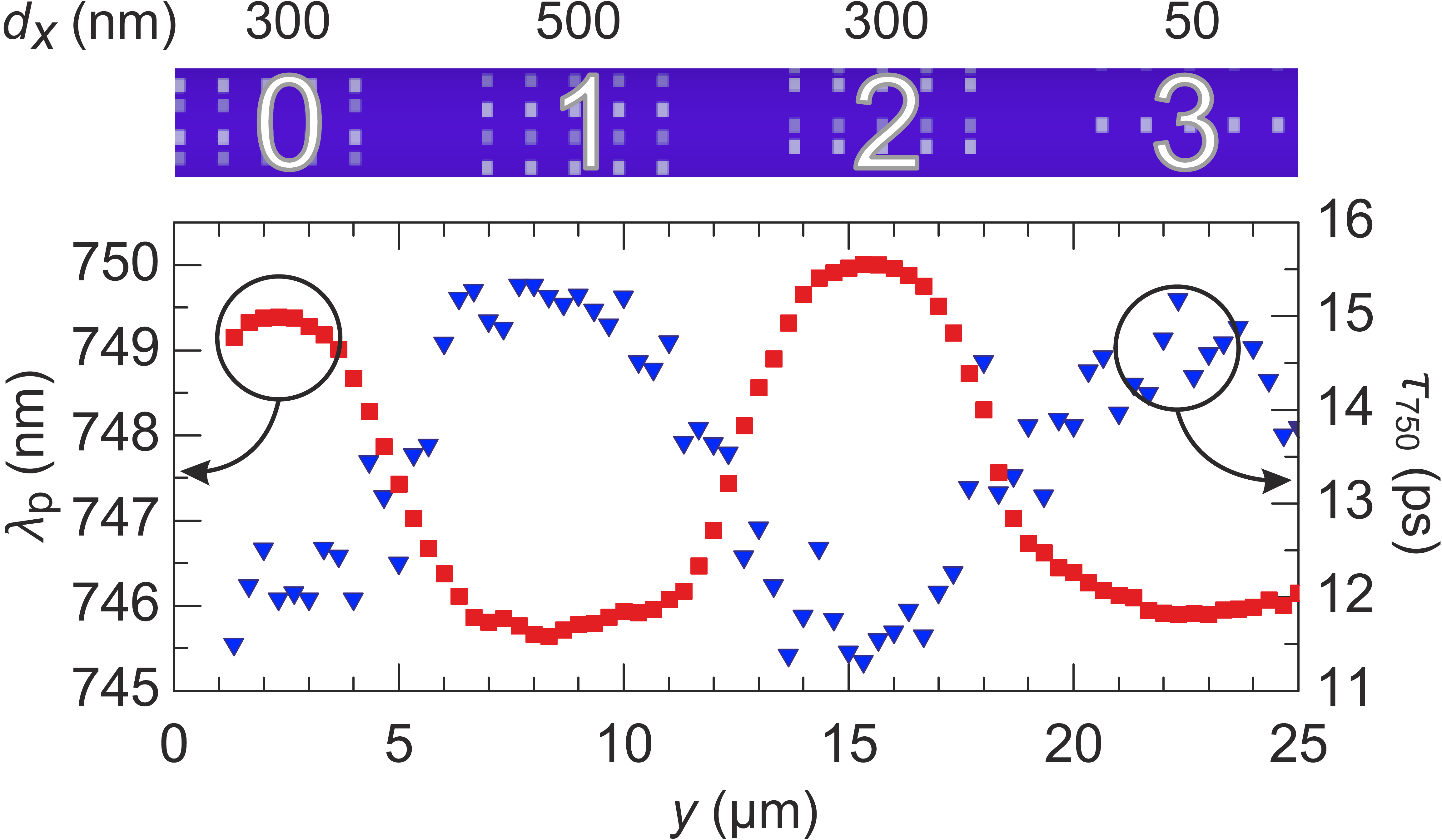}
	\caption{Profiles along the $y$-direction of the spectral position of the maximum PL intensity $\lambda_p$ and of the QW PL lifetime at an emission wavelength of \SI{750}{\nm} $\tau_\text{750}$. Data points are taken from the scan area framed by the dashed box in Fig.\,\ref{fig:ext_sim_data}\,(b), averaged in $x$-direction.}
	\label{fig:profiles}
\end{figure}

We further analyzed the time-resolved measurements by evaluating the spectrally resolved PL decay curves for the $749$--$\SI{751}{\nm}$ wavelength range. The decay curves are fitted to a monoexponential function and the resulting lifetimes are denoted as $\tau_\text{750}$. We concentrate on the scan area framed in the dashed box in Fig.\,\ref{fig:ext_sim_data}\,(b) which includes the arrays $0$ to $3$ and thus cuboid distances $d_x$ of \SI{300}{\nm}, \SI{500}{\nm}, \SI{300}{\nm} and \SI{50}{\nm}. The data points for $\lambda_p$ and $\tau_\text{750}$ are averaged in $x$-direction and are plotted as profiles in $y$-direction in Fig.\,\ref{fig:profiles}. The spectral position of the maximum PL intensity $\lambda_p$ clearly shifts to \SI{749.5}{\nm} and \SI{750}{\nm} in array $0$ and $2$, respectively. The spectral redshift is accompanied by a decrease in lifetime $\tau_\text{750}$ to \SI{12}{\ps} and \SI{11}{\ps}, respectively. The subtle differences in $\lambda_p$ and $\tau_\text{750}$ between array 0 and 2, which have nominally the same geometries, are primarily caused by small irregularities of the cuboids themselves. According to the time-resolved measurements, the nanoantenna with $d_x=\SI{300}{\nm}$ selectively enhances the transition corresponding to a wavelength range from \SI{749}{\nm} to \SI{751}{\nm}. The enhancement of that spectral part induces an rearrangement of the quantum well spectra, i.e., the transitions corresponding to that wavelength range are highly preferred which leads to the observed redshift of the overall PL emission.

\begin{figure*}
		\includegraphics[scale=0.175]{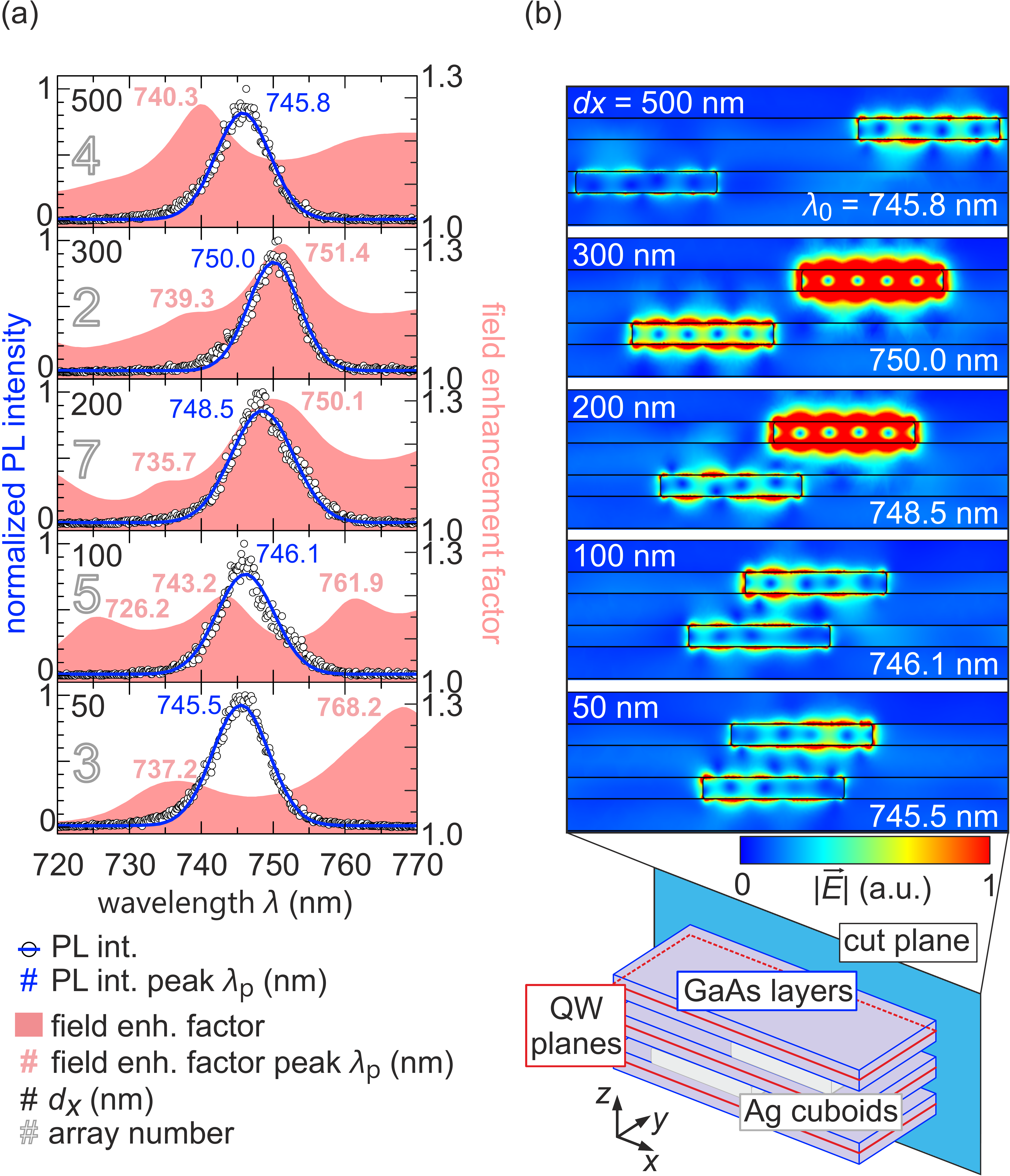}
	\caption{(a) QW emission spectra (black circles), gaussian fit curves (solid lines) and simulated field enhancement factors (red filled curves) for cuboid's distances of \SI{500}{\nm}, \SI{300}{\nm}, \SI{200}{\nm}, \SI{100}{\nm} and \SI{50}{\nm} (top to bottom). (b) Distribution of the absolute electric field $|\vec{E}|$ for the same set of distances $d_x$. The field distribution is evaluated at an excitation wavelength $\lambda_0$ corresponding to the respective measured PL emission peak $\lambda_p$. Cross sections are taken from a cut plane as shown in the bottom.}
	\label{fig:sim}
\end{figure*}

We carried out finite-element simulations with \textsc{comsol multiphysics} to calculate the field enhancement factor in the plane of the QWs and the overall distribution of the absolute electric field $|\vec{E}|$. The simulations allow to identify both the spatial and spectral overlap of emitter and plasmonic antenna and are performed for the Ag geometries corresponding to the experimentally realized distances, i.e., $d_x=\SI{50}{\nm}$ (array $3$), $d_x=\SI{100}{\nm}$ (array $5$), $d_x=\SI{200}{\nm}$ (array $7$), $d_x=\SI{300}{\nm}$ (array $0$, $2$ and $8$), and $d_x=\SI{500}{\nm}$ (array $1$ and $4$). For simplicity we neglect the curvature of the system and assume a geometry which consists of two Ag cuboids each with size \SI[product-units = brackets]{250x250x30}{\nm^3} that are sandwiched between three $45$-nm thick GaAs slabs as depicted in the sketch at the bottom of Fig.\,\ref{fig:sim}\,(b). The semiconductor heterostructure is approximated as a homogenous GaAs layer and the values for the complex permittivities of GaAs and Ag are taken from Ref.\cite{Aspnes1983} and Ref. \cite{Johnson1972}, respectively. The system is excited by a plane wave which is polarized in direction parallel to the cuboids' displacement and impinging from top onto the structure.

For calculating the field enhancement factors the integrated absolute electric field in the planes of the QW is normalized with respect to the integrated electric field in corresponding planes of the layered system but without Ag cuboids. The field enhancement spectra are plotted as red filled curves in Fig.\,\ref{fig:sim}\,(a) and peaks in the spectra indicate resonant wavelengths with increased field intensities corresponding to localized surface plasmon modes. For $d_x=\SI{50}{\nm}$ the simulated spectra exhibit two peaks at \SI{737.2}{\nm} and \SI{768.2}{\nm}. For a distance of $d_x=\SI{100}{\nm}$ three peaks at \SI{726.2}{\nm}, \SI{743.2}{\nm}, and \SI{761.9}{\nm} are observable. For $d_x=\SI{200}{\nm}$ a broad peak appears at \SI{750.1}{\nm} which gets narrower and slightly red-shifted to \SI{751.4}{\nm} for $d_x=\SI{300}{\nm}$. Here, in addition, shoulder peaks at \SI{735.7}{\nm} and \SI{739.3}{\nm} are visible, respectively. For $d_x=\SI{500}{\nm}$ the main spectral peak shifts to the blue to \SI{740.3}{\nm}. To compare with the experimental data, normalized QW emission spectra (black circles) and corresponding Gaussian fit curves (blue solid lines) taken from representative pixels marked in Fig.\,\ref{fig:ext_sim_data}\,(b) are also shown in Fig.\,\ref{fig:sim}\,(a). It becomes obvious that there is a matching between the calculated field enhancement and the measured PL peaks only for $d_x=\SI{200}{\nm}$ and $d_x=\SI{300}{\nm}$. In these cases, the measured PL spectra are red-shifted compared to the other spectra. This observation strongly suggests that the redshift is due to a coupling of the QW emission to the localized surface plasmon modes of the nanoantenna. For the other nanoantenna geometries, the plasmonic resonances are detuned and the QW emission spectrum remains unaltered. We further analyzed the distribution of the absolute electric field $|\vec{E}|$ inside the Ag cuboid nanoantenna to characterize the involved plasmon modes. For this purpose we evaluate the distribution at an excitation wavelength $\lambda_0$ which corresponds to the respective measured PL emission peak wavelength $\lambda_p$. Figure \ref{fig:sim}\,(b) shows the cross-sectional plane of the distribution of the absolute electric field $|\vec{E}|$ taken from the cut plane as depicted in the bottom. The different field distributions of both cuboids in each of the panel of Fig.\,\ref{fig:sim}\,(b) are caused by retardation effects, i.e., the plane wave impinges the cuboid system from above and thus excites the upper cuboid first. For $d_x=\SI{300}{\nm}$, a symmetrical high-order plasmonic mode emerges in both cuboids. Remarkably, the field intensities are distributed around each cuboid and are not condensed in a high intensity hot spot in-between. This mode distribution is also apparent in the upper cuboid for $d_x=\SI{200}{\nm}$, however the field distribution of the other cuboid is altered and disturbed. For distances $d_x$ of \SI{500}{\nm}, \SI{100}{\nm} and \SI{50}{\nm}, no symmetrically distributed plasmonic mode is visible. The resonant coupling of the QW emission to the nanoantenna with cuboid distance of $d_x=\SI{300}{\nm}$ is therefore associated with an excitation of a high-order plasmonic mode emerging in both cuboids.

We want to emphasize that the presented hybrid emitter/nanoantenna design is not limited to GaAs QWs and Ag cuboids. The rolling-up fabrication allows to incorporate other quantum emitters into the strained bilayer system, like self-assembled quantum dots \cite{Mendach2006,Kipp2006,Li2009}. Furthermore, it is not limited to the InAlGaAs material system; other systems with larger band gaps like AlInP led to rolled-up structures \cite{Strelow2012} that allow for plasmonic coupling to emitters over essentially the whole visible spectral range. Besides incorporating quantum emitters directly into the strained layer system, it is also possible to utilize the strained bilayer just as a host onto which other emitters and the plasmonic nanostructures are successively deposited before the rolling process. These emitters might be thin films of colloidal nanocrystals or organic light-emitting molecules \cite{Kietzmann2015}. Very promising, these thin films can also be lithographically structured into stripes \cite{Kietzmann2015} or squares before rolling. The separate preparation of emitter and metal nanostructures onto the strained bilayer enables remarkable emitter/nanoantenna designs. For example, rolling-up these emitters together with predefined Ag cuboids as used in the current work can lead to a close stacking of emitters in-between metal structures that can only hardly be realized in conventional subsequent lateral lithographic patterning alone. As for the metal part, the EB lithography approach facilitates also other antenna designs like particle dimers or bow ties. Smaller metal particles with fundamental dipole resonances in the visible spectrum allow to further modify the LSP resonance and to enhance the on-resonance near-field intensity. Besides that, the tubular geometry allows to exploit plasmon-photon whispering gallery modes \cite{Rottler2013,Yin2016} for an even more complex tailoring of the light-matter interaction. The fabrication ansatz therefore represents an important step for the realization of quantum light sources with tailor-made optical properties. 

In conclusion we report on a novel fabrication approach for a tunable plasmonic nanoantenna. We integrated Ag cuboids into a rolled-up semiconductor microtube such that two of these cuboids are stacked at a certain distance with the QW in between. The lateral distance $d_x$ of the cuboids is deliberately changed to tune the antenna's resonance. By means of spatially, spectrally and temporally resolved PL measurements we show a coupling of the quantum-well emission to the nanoantenna for a cuboid distance of $d_x=\SI{300}{\nm}$. We observe a redshift which is accompanied by a lifetime reduction of the the quantum-well emission. Finite-element simulations reveal that an excitation of high-order plasmonic modes inside the cuboids accounts for the resonant coupling.

\begin{acknowledgement}
We acknowledge financial support from Deutsche Forschungsgemeinschaft grant No.~ME 3600/1-1. T.~Kipp also acknowledges funding from the European Union's Horizon 2020 research and innovation programme under the Marie Sk\l odowska-Curie grant agreement No.~656598.  
\end{acknowledgement}

\bibliography{cuboid_paper_v3}

\end{document}